\documentclass[twocolumn,prb,showpacs,showkeys,amsmath,amssymb,floatfix]{revtex4-1}
\usepackage{graphicx} 
\usepackage{dcolumn}  
\usepackage{bm}       
\usepackage{natbib}

\usepackage{psfrag}   
\usepackage{pifont}
\usepackage{gensymb,wasysym,mathrsfs,bbm}
\usepackage{tensind}  
\usepackage{hyperref}

\usepackage{color}
\usepackage{cases}

\def\setR{\mathbbm{R}}

\begin{document}
\newcommand{\VertSpace}{\vspace{2mm}}
\tensordelimiter{?}

\preprint{unknown}
\title{Twin Paradox in de Sitter Spacetime}

\author{Sebastian Boblest}
\email{sebastian.boblest@itp1.uni-stuttgart.de}
\affiliation{
   Universit\"at Stuttgart, 1. Institut f\"ur Theoretische Physik\\
   Pfaffenwaldring 57 // IV,  70569 Stuttgart, Germany
}

\author{Thomas M{\"u}ller}
\email{Thomas.Mueller@vis.uni-stuttgart.de}
\affiliation{
  Visualisierungsinstitut der Universit\"at Stuttgart (VISUS)\\
  Allmandring 19,
  70569 Stuttgart, Germany
}

\author{G\"unter Wunner}
\email{guenter.wunner@itp1.uni-stuttgart.de}
\affiliation{
   Universit\"at Stuttgart, 1. Institut f\"ur Theoretische Physik\\
   Pfaffenwaldring 57 // IV,  70569 Stuttgart, Germany{}
}

\begin{abstract}
 The "twin paradox" of special relativity offers the possibility to make 
interstellar flights
within a lifetime. For very long journeys with velocities close to the 
speed of light, however, 
we have to take into account the expansion of the universe. Inspired by 
the work of Rindler on
hyperbolic motion in curved spacetime, we study the worldline of a 
uniformly accelerated observer 
in de Sitter spacetime and the communication between the traveling 
observer and an observer at rest.
\end{abstract}

%


\maketitle

\section{Introduction}\label{sec:intro}
The ``twin paradox'' is one of the most discussed problems in special relativity\cite{Rindler2006, French1971}. The term ``paradox'', however, is quite misleading, as all ramifications of the situation considered are well understood nowadays. 
Nevertheless, we use this term to sketch the basic situation of this paper. While one of the twins, we will call him Eric, stays on Earth, the other twin, Tina, undertakes a relativistic trip to some distant location in the universe and immediately returns home. In a previous work, M\"uller et al. \cite{Mueller2006} discussed this situation in flat Minkowski space for a uniformly accelerating twin. There, the journey is separated into four stages of equal time lengths. In the first two stages, Tina accelerates and decelerates again until she reaches her destination. In the last two stages, she returns in exactly the same manner. As long as Tina visits only a nearby star, the special relativistic treatment is a sufficient approximation. But, if a trip to some far distant galaxies shall be considered, the expansion of spacetime must be taken into account. 

The aim of this paper is to discuss the twin paradox situation for a uniformly accelerating twin in de Sitter spacetime. Although our own universe is not a de Sitter spacetime, nevertheless this will reveal the new aspects of the twin paradox that come about in an expanding universe. 
As pointed out by Rindler \cite{Rindler1960} it has the didactive merit that in it all the necessary integrations can be performed in terms of elementary functions. We extend the calculations by Rindler to incorporate not only the acceleration in one direction but also the complete journey. For very long acceleration times or very large Hubble constants, the temporal course of the journey will then differ significantly from the flat Minkowskian situation. Besides the worldline of the twin, we also study the influence of the expanding universe and the accelerated motion on the communication between both twins. 

Detailed discussions of the structure of the de Sitter space\cite{deSitter1917a,deSitter1917b} can be found in Hawking and Ellis\cite{hawkingellis}, Schmidt\cite{schmidt1993}, or Spradlin et al.\cite{Spradlin2001} There are also some recent articles concerning accelerated motion in de Sitter space. Bi\v{c}\'ak and Krtou\v{s}\cite{Bicak2001}, for example, study accelerated sources in de Sitter space and give a list of several coordinate representations. Podolsk\'y and Griffiths\cite{Podolsky2000} discuss uniformly accelerating black holes in a de Sitter universe. From an other point of view, Doughty\cite{doughty1981} discusses the necessary acceleration in Schwarzschild spacetime to keep a fixed distance to the black hole horizon.

Some recent publications on uniform acceleration within special relativity are, for example, Semay\cite{semay2007}, who presents a uniformly accelerated observer within a Penrose-Carter diagram, or Flores\cite{flores2008}, who is concerned with the communication between accelerated observers.

The structure of this paper is as follows. In Sec.~\ref{sec:HypFlat}, we briefly summarize the twin paradox journey in flat Minkowski space. In Sec.~\ref{sec:Prelim}, we introduce the form of the de Sitter metric we use in this work and recapitulate some properties needed for our discussion. In Sec.~\ref{sec:HypMotion}, we derive the worldline of a twin paradox journey in de Sitter spacetime and discuss the differences to the journey in Minkowski space. In Sec.~\ref{sec:Communication}, we scrutinize the influence of the expansion onto the communication between both twins.

\section{Twin Paradox in Flat Space}\label{sec:HypFlat}
\begin{figure}[htb]
 \includegraphics[width=3.5in]{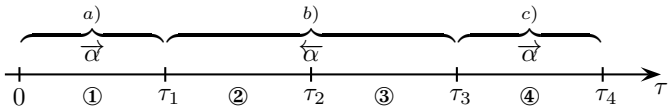}
 \caption{Branches, acceleration directions, and proper times with respect to Tina. In Minkowski space, we have $\tau_n=n \tau_1$ for $n=\{2,3,4\}$.}
 \label{fig:branches}
\end{figure}

In flat Minkowski space, the twin paradox journey works as follows. While Eric stays on Earth, Tina starts her trip with zero velocity and moves with uniform acceleration $\alpha$. After proper time $\tau_1$ with respect to her own clock, she decelerates again with $\alpha$ until she comes to rest at proper time $2\tau_1$. Then, she immediately returns to Earth with the same procedure and reaches her brother at $4\tau_1$. Here, and for the rest of this article, we denote these four stages with the circled numbers \ding{172}--\ding{175}. Because the accelerations in the stages~\ding{173} and \ding{174} point in the same direction, each worldline is composed of three branches. Branch $a)$ describes stage~\ding{172}, stages~\ding{173} and \ding{174} are combined into branch $b)$, and stage~\ding{175} is represented by branch $c)$, see Fig.~\ref{fig:branches}.

Hence, Tina's worldline with respect to Minkowski spacetime, $ds^2=-c^2dt^2+dr^2+r^2(d\vartheta^2+\sin^2\vartheta\,d\varphi^2)$, reads $(t=t_{\text{\tiny flat}}(\tau), r=r_{\text{\tiny flat}}(\tau), \vartheta=\pi/2, \varphi=0)$ with
\begin{subnumcases}{\label{eq:tMink}t_{\text{\tiny flat}}(\tau) = \dfrac{c}{\alpha}}
                    \textstyle \sinh\left(\frac{\alpha}{c}\tau\right) \label{eq:tMink1}\\
		    \textstyle \sinh\left[\frac{\alpha}{c}\left(\tau - 2\tau_1\right)\right] + 2\sinh\left(\frac{\alpha}{c}\tau_1\right)\label{eq:tMink2} \\
		    \textstyle \sinh\left[\frac{\alpha}{c}\left(\tau - 4\tau_1\right)\right] + 4\sinh\left(\frac{\alpha}{c}\tau_1\right)\label{eq:tMink3}
\end{subnumcases}
and
\begin{subnumcases}{\label{eq:xMink}\hspace*{-0.6cm} r_{\text{\tiny flat}}(\tau) = \dfrac{c^2}{\alpha}}
                     \textstyle \cosh\left(\frac{\alpha}{c}\tau\right)-1 \label{eq:xMink1} \\
		     \textstyle 2\cosh\left(\frac{\alpha}{c}\tau_1\right) -\cosh\left[\frac{\alpha}{c}\left(\tau - 2\tau_1\right)\right]-1\label{eq:xMink2} \\
		     \textstyle \cosh\left[\frac{\alpha}{c}\left(\tau - 4\tau_1\right)\right] - 1\label{eq:xMink3}.
\end{subnumcases}
A derivation of this worldline can be found in App.~\ref{appsec:hypmotionflat} or in M{\"u}ller et al.\cite{Mueller2006}

\begin{figure}[htb]
 \includegraphics[width=3.4in]{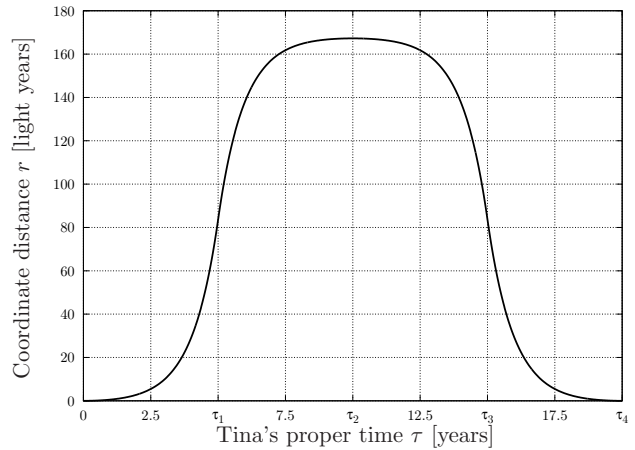}
 \caption{Radial coordinate distance $r$ for a round trip with $4\tau_1=20\,\text{y}$. Tina's maximum distance to Eric is $r_{\text{\tiny max}}\approx 167.19\,\text{ly}$.}
 \label{fig:XMink}
\end{figure}

As an example, Fig.~\ref{fig:XMink} shows Tina's worldline with respect to her proper time $\tau$. In each stage, she accelerates or decelerates with $\alpha=9.81\,\text{m}/\text{s}^2\approx 1.0326\,\text{ly}/\text{y}^2$ for a time $\tau_1=5\,\text{y}$. Thus, her journey lasts $4\tau_1=20\,\text{y}$. On Earth, however, $t(4\tau_1)\approx 338.36\,\text{y}$ pass by, as can easily be calculated from Eq.~(\ref{eq:tMink3}). The maximum coordinate distance, $r_{\text{\tiny max}}$, Tina can reach with this procedure, is given by Eq.~(\ref{eq:xMink2}), setting $\tau=2\tau_1$.

\begin{figure}[htb]
 \includegraphics[width=3.4in]{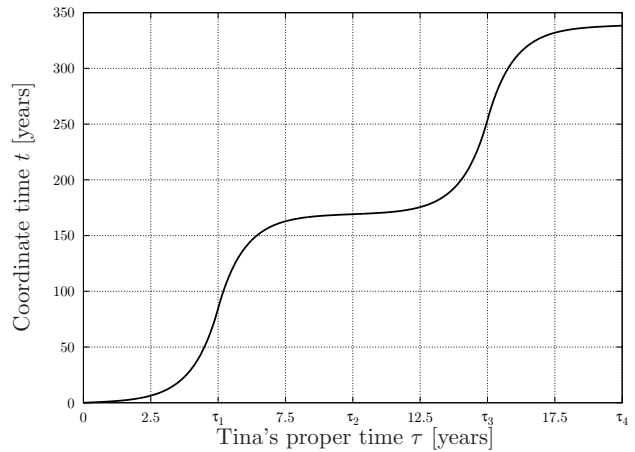}
 \caption{Coordinate time $t$ for the round trip of Fig.~\ref{fig:XMink}.}
 \label{fig:TMink}
\end{figure}

In Fig.~\ref{fig:TMink}, the coordinate time $t(\tau)$ is shown for this trip. Note the pointwise symmetry of $t(\tau)$ around $2\tau_1$. This feature is lost in de Sitter space.

\section{Basic properties of de Sitter space} \label{sec:Prelim}
The de Sitter spacetime can be described by several coordinate systems, see e.g. Bi\v{c}\'ak and Krtou\v{s}~\cite{Bicak2001}. In conformal Einstein (CE) coordinates, the de Sitter metric reads
\begin{equation}
  ds^2 = \frac{\kappa^2}{\sin^2\eta}\left[-d\eta^2+d\chi^2+\sin^2\chi\,d\omega^2\right],
  \label{eq:confEinstMetric}
\end{equation}
where $d\omega^2=d\vartheta^2+\sin^2\vartheta\,d\varphi^2$ is the spherical surface element, and $\kappa=c/H$ is given by the speed of light $c$ and the Hubble constant $H$. Here, the coordinates $\eta$ and $\chi$ are restricted to $\eta\in(0,\pi)$ and $\chi\in[-\pi,\pi]$, where $\chi=-\pi$ and $\chi=\pi$ are identified. 

To follow the calculations for accelerated motion by Rindler\cite{Rindler1960}, we mainly use Lema\^{i}tre-Robertson\cite{Lemaitre1925,Robertson1928} (LR) coordinates with line element
\begin{equation}
 ds^2 = -c^2dt^2 + e^{2Ht}\left(dr^2 + r^2d\omega^2\right).
\label{eq:metric}
\end{equation}
Here, $(t,r)\in\setR$, however, we identify points $(r<0,\vartheta,\varphi)$ with $(r>0,\pi-\vartheta,\varphi-\pi)$. Both coordinate systems are related via the following transformation equations
\begin{subequations}
 \begin{alignat}{5}
    \label{eq:coordTransf1} T &= \frac{\kappa\sin\eta}{\cos\chi+\cos\eta}, &\quad r &= \frac{\kappa\sin\chi}{\cos\chi+\cos\eta},\quad\text{or}\\
    \label{eq:coordTransf2} \eta &= \arctan\frac{2T\kappa}{\kappa^2-T^2+r^2},&\quad \chi &= \arctan\frac{2r\kappa}{\kappa^2+T^2-r^2},
 \end{alignat}
\end{subequations}
where $T=\kappa\exp(-ct/\kappa)$ or $ct=\kappa\ln(\kappa/T)$, respectively. While (\ref{eq:coordTransf1}) is unique, we have to take care of the coordinate domains in (\ref{eq:coordTransf2}). In that case, if $\kappa^2-T^2+r^2<0$, we have to map $\eta\rightarrow \eta+\pi$. On the other hand, if $\kappa^2+T^2-r^2<0$, we have to consider the sign of $r$. If $r>0$, then $\chi\rightarrow \chi+\pi$, otherwise $\chi\rightarrow \chi-\pi$.

\begin{figure}[htb]
 \includegraphics[width=3.5in]{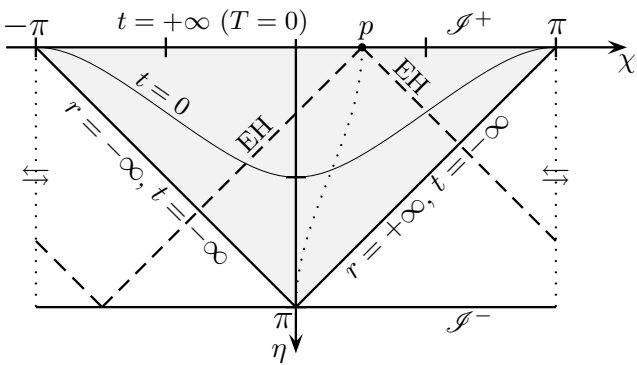}
 \caption{Penrose diagram of de Sitter spacetime in conformal Einstein (CE) coordinates $\chi$ and $\eta$. The event horizon for a static observer $\mathcal{O}$ (dotted line) at $p$ is indicated by dashed lines (EH). Note that $t=-\infty$ corresponds to $T=+\infty$. The left-right-arrows indicate that the `lines' $\chi=-\pi$ and $\chi=\pi$ are identified. The gray-shaded triangle corresponds to the LR coordinate domain.}
 \label{fig:penroseDiagram}
\end{figure}

The advantage of the CE coordinates is that in the Penrose diagram, Fig.~\ref{fig:penroseDiagram}, radial light rays $(d\vartheta=d\varphi=0)$ are represented by straight lines with $\pm 45\degree$ slope. All past-directed light rays end at $\mathscr{I}^{-}$ $(\eta=\pi)$, whereas all future-directed ones end at $\mathscr{I}^{+}$  $(\eta=0)$. In contrast to the LR coordinates, the CE coordinates cover the whole spacetime. But, for our purpose, the LR coordinate domain is adequate.

The hypersurface $t=0$ in CE coordinates is independent of $\kappa$:
\begin{equation}
 \eta(\chi) = \arctan\left[\xi_{+}(\chi),\xi_{-}(\chi)\right]
\end{equation}
with $\xi_{\pm}(\chi)=\frac{1}{2}(\mp\cos\chi+\sqrt{2-\cos^2\chi})$.

The worldline of a static observer $\mathcal{O}$ with respect to the LR coordinates, $r=\text{const}$, is represented by the dotted line. The backward light cone of $\mathcal{O}$'s ``final'' point $p$ defines his event horizon (EH). 

\begin{figure}[htb]
 \includegraphics[width=3.5in]{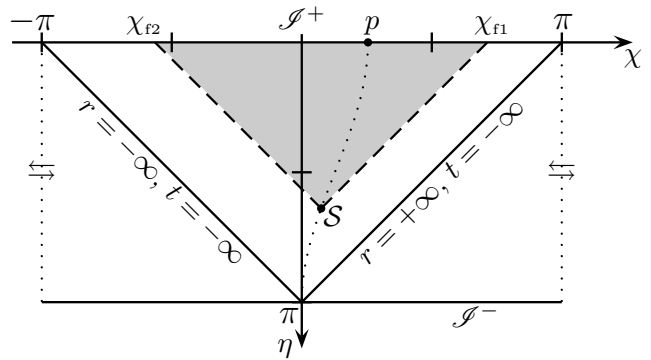}
 \caption{The dashed lines represent the forward light cone of $\mathcal{S}$ with coordinates $(t_{\tiny\mathcal{S}}=-0.5,r_{\tiny\mathcal{S}}\approx 0.423)$ or $(\eta_{\tiny\mathcal{S}}\approx 1.482,\chi_{\tiny\mathcal{S}}\approx 0.435)$, respectively. As in Fig.~\ref{fig:penroseDiagram}, the dotted line is the worldline of the static observer $\mathcal{O}$.}
 \label{fig:forwardLightcone}
\end{figure}

For our twin paradox journey, we are interested in the radial domain that can be reached by Tina when she starts at point $\mathcal{S}$ with coordinates $(t_{\tiny\mathcal{S}},r_{\tiny\mathcal{S}})$ or $(\eta_{\tiny\mathcal{S}},\chi_{\tiny\mathcal{S}})$, respectively. The radial domain follows from the forward light cone of $\mathcal{S}$, see Fig.~\ref{fig:forwardLightcone}

The critical points, where the forward light cone intersects $\eta=0$, read $\chi_{\text{\tiny f1}} = \chi_{\tiny\mathcal{S}}+\eta_{\tiny\mathcal{S}}$ and $\chi_{\text{\tiny f2}} = \chi_{\tiny\mathcal{S}}-\eta_{\tiny\mathcal{S}}$. The corresponding LR radial coordinates are given by
\begin{equation}
 r_{\text{\tiny f1}} = \frac{\kappa\sin(\chi_{\tiny\mathcal{S}}+\eta_{\tiny\mathcal{S}})}{\cos(\chi_{\tiny\mathcal{S}}+\eta_{\tiny\mathcal{S}})+1}\quad\text{and}\quad r_{\text{\tiny f2}} = \frac{\kappa\sin(\chi_{\tiny\mathcal{S}}-\eta_{\tiny\mathcal{S}})}{\cos(\chi_{\tiny\mathcal{S}}-\eta_{\tiny\mathcal{S}})+1}.
 \label{eq:critForwardPoints}
\end{equation}
If $\chi_{\tiny\mathcal{S}}=0$, Eq.~(\ref{eq:critForwardPoints}) can be simplified to
\begin{equation}
  r_{\text{\tiny f1,f2}} = \pm \kappa e^{-ct_{\tiny\mathcal{S}}/\kappa}.
\label{eq:EventHorizon}
\end{equation}
Hence, if Tina starts at $\mathcal{S}$, she can only move within the gray-shaded region.

\begin{figure}[htb]
 \includegraphics[width=3.5in]{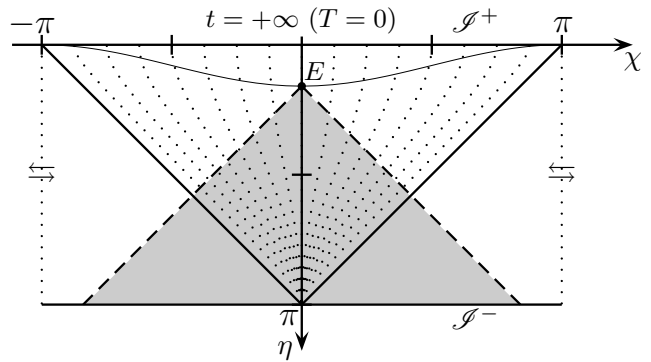}
 \caption{The dashed lines represent Eric's (E) backward light cone at the current observation time $\eta_{\text{\tiny E}}=0.5$ or $t_{\text{\tiny E}}\approx 1.365$, respectively. The dotted lines represent static observers, $r=\text{const}$, with respect to LR coordinates. The thin solid line indicates the hypersurface $t=t_{\text{\tiny E}}$.}
 \label{eq:ericLightcone}
\end{figure}

Eric's worldline is represented either by $(t,r_{\text{\tiny E}}=0)$ or by $(\eta,\chi_{\text{\tiny E}}=0)$, where $\eta=\arctan\{2\exp(-ct/\kappa)/[1-\exp(-2ct/\kappa)]\}$. He can only observe events that lie inside his backward light cone (gray-shaded region in Fig.~\ref{eq:ericLightcone}).

The event horizon of Eric in CE coordinates is defined by $\eta=\pm\chi$, when he is located at $(\eta_{\text{\tiny E}}=0,\chi_{\text{\tiny E}}=0)$. In LR coordinates the event horizon simplifies to $r=\kappa\exp(-ct/\kappa)$.

In an expanding spacetime, the difference of coordinates of two points has little meaning as a measure of distance. 
Therefore, we follow Rindler\cite{Rindler1960} and introduce the proper radial distance
\begin{equation}
 l = e^{Ht}r,
\label{eq: propdist}
\end{equation}
between Eric $(r_{\text{\tiny E}}=0)$ and a point $Q$ located at coordinate distance $r$. The proper distance can be interpreted as the result of a distance measurement, where infinitely many stationary observers between Eric and $Q$ sum up the distances they measure between each other at a given coordinate time\cite{Rindler2006}. Obviously, if $t=0$, proper distance equals coordinate distance. 

\section{Twin Paradox in de Sitter space }\label{sec:HypMotion}

\subsection{Derivation of the Worldline}\label{sec:WLcalc}
In general relativity, the worldline $x^{\mu}(\tau)$ of an individual moving with constant proper acceleration $\alpha$ with respect to its local reference frame follows from
\begin{equation}
  g_{\mu\nu}a^{\mu}a^{\nu} = \alpha^2
  \label{eq:hypmotion}
\end{equation}
with the four-acceleration
\begin{equation}
 a^{\mu} = \frac{d^{2}x^{\mu}}{d\tau^{2}} + \Gamma_{\alpha\beta}^{\mu}\frac{dx^{\alpha}}{d\tau}\frac{dx^{\beta}}{d\tau}.
 \label{eq:amu}
\end{equation}
Here, $\tau$ is the proper time of the individual and $\Gamma_{\alpha\beta}^{\mu}$ are the Christoffel symbols of the corresponding spacetime. In addition, the constraint equation $g_{\mu\nu}u^{\mu}u^{\nu} = -c^2$ for the four-velocity $u^{\mu}=dx^{\mu}/d\tau$ must be fulfilled.

For the de Sitter space, the worldline of an individual who accelerates away from Earth for all times was determined by Rindler\cite{Rindler1960}. To study a round trip as in Minkowski space, Eq.~(\ref{eq:hypmotion}) has to be solved for all three branches of this journey separately. At the beginning, Tina starts from Earth with zero velocity and moves with constant acceleration $\alpha$ for some time $\tau_1$. Thus, at Tina's proper time $\tau=0$, we have
\begin{equation}
 t=0,\quad r=0,\quad t'=1,\quad r'=0.
\end{equation}
Note that here and in the following, a prime denotes differentiation with respect to $\tau$. 

Contrary to flat space, we allow different durations for the four stages of the journey for reasons we will present later. Taking into account that $t'(\tau)$ is continuous everywhere, we obtain
\begin{subnumcases}{\label{eq:tscomplete}t'(\tau) = q}
 \frac{Se^{2q\tau}+D}{\Psi(\tau)} \label{eq:tscomplete1} \\ 
 \frac{\frac{D^2}{S}A_{\tau_1}^2e^{2q\tau}+D}{\Psi_A(\tau)} \label{eq:tscomplete2} \\ 
 \frac{\frac{D^2}{S}B_{\tau_3}^2e^{2q\tau}+D}{\Psi_B(\tau)} \label{eq:tscomplete3},
\end{subnumcases}
where 
\begin{equation}
 q = \sqrt{\left(\frac{\alpha}{c}\right)^2 + H^2},\quad S = q+H, \quad D=q-H
\label{eq:RindlerqHSD}
\end{equation}
and
\begin{subequations}
 \begin{align}
   \Psi(\tau) &= HSe^{2q\tau} + 2SDe^{q\tau} - HD,\label{eq:RindlerPsi}\\
    A_{\tau_1} &= \frac{\Omega_{\tau_1}S - \frac{\alpha}{c}\left[\Omega_{\tau_1}-2\Psi(\tau_1)\right]}{\Omega_{\tau_1}D + \frac{\alpha}{c}\left[\Omega_{\tau_1}-2\Psi(\tau_1)\right]}e^{-q\tau_1},\label{eq:Atau1}\displaybreak[0]\\
    B_{\tau_3} &= \frac{\Omega_{\tau_3}S + \frac{\alpha}{c}\left[\Omega_{\tau_3}-2\Psi_A(\tau_3)\right]}{\Omega_{\tau_3}D - \frac{\alpha}{c}\left[\Omega_{\tau_3}-2\Psi_A(\tau_3)\right]}e^{-q\tau_3},\label{eq:Btau2}\displaybreak[0]\\
     \Omega_{\tau_1} &= \textstyle S\left(S+\frac{\alpha}{c}\right)e^{2q\tau_1} \label{eq:omega1} \\
	\nonumber  &\quad\textstyle + 2\left(SD-\frac{\alpha}{c}H\right)e^{q\tau_1} + D\left(D-\frac{\alpha}{c}\right),\displaybreak[0]\\
    \Omega_{\tau_3} &= \textstyle\frac{D^2}{S}A_{\tau_1}^2\left(S-\frac{\alpha}{c}\right)e^{2q\tau_3} \\
	\nonumber	&\quad \textstyle +2\frac{D}{S}A_{\tau_1}\left(SD+\frac{\alpha}{c}H\right)e^{q\tau_3} + D\left(D+\frac{\alpha}{c}\right),\displaybreak[0]\\
   \Psi_{A}(\tau) &= \textstyle \frac{A_{\tau_1}^2D^2}{S}He^{2q\tau} +2A_{\tau_1}D^2e^{q\tau} -HD,\\
   \Psi_{B}(\tau) &= \textstyle \frac{B_{\tau_3}^2D^2}{S}He^{2q\tau} +2B_{\tau_3}D^2e^{q\tau} -HD.
 \end{align}
\end{subequations}
(Here, the branches $a)$ to $g)$ do not denote the different branches of the worldline!)

By definition, we have
\begin{equation}
  t'(\tau)= \gamma(\tau) = \frac{1}{\sqrt{1-\beta(\tau)^2}},
\label{eq:defgamma}
\end{equation}
with the Lorentz-factor $\gamma$ of special relativity and $\beta = v/c$, where $v$ is Tina's velocity with respect to a local observer at rest at her current position.
With Eqs.~(\ref{eq:tscomplete}) and (\ref{eq:defgamma}) we obtain
\begin{subnumcases}{\label{eq:betacomplete}\beta(\tau) = \frac{\alpha}{cq}}
 \frac{\left(e^{q\tau}-1\right)\left(Se^{q\tau}+D\right)}{Se^{2q\tau}+D} \label{eq:betacomplete1} \\ 
 -\frac{\left(A_{\tau_1}\frac{D}{S}e^{q\tau}-1\right)\left(DA_{\tau_1}e^{q\tau}+D\right)}{\frac{D^2A_{\tau_1}}{S}e^{2q\tau}+D} \label{eq:betacomplete2} \\ 
 \frac{\left(B_{\tau_2}\frac{D}{S}e^{q\tau}-1\right)\left(DB_{\tau_2}e^{q\tau}+D\right)}{\frac{D^2B_{\tau_2}}{S}e^{2q\tau}+D} \label{eq:betacomplete3},
\end{subnumcases}
for her velocity during her trip.
\\
An important aspect in the following discussion is the maximum possible velocity on a journey.  
The maximum velocity is reached if Tina accelerates for all times. Using Eqs.~(\ref{eq:tscomplete1}) and (\ref{eq:betacomplete1}) we obtain
\begin{equation}
 \gamma_{\infty}=\lim_{\tau\rightarrow\infty}\gamma(\tau) = \frac{q}{H},\quad  \beta_{\infty}=\lim_{\tau\rightarrow\infty}\beta(\tau) = \frac{1}{q}\frac{\alpha}{c}.
\label{eq:maxgamma}
\end{equation}
Because $q>\alpha/c$, Tina's velocity asymptotically reaches some value smaller than 1, contrary to flat space, where $\beta\rightarrow 1$ 
for infinitely long trips. 

Integrating Eq.~(\ref{eq:tscomplete}) over $\tau$, and adjusting the constants of integration such that $t(\tau)$ is continuous and $t(0)=0$, yields 
\begin{subnumcases}{\label{eq:tcomplete}t(\tau) = \frac{1}{H}}
   \ln\left[\frac{\Psi(\tau)}{2q^2 e^{q\tau}}\right]\label{eq:tcomplete1} \\
   \ln\left[\frac{\Psi_A(\tau)}{2q^2 e^{q\tau}}\frac{\Psi(\tau_{1})}{\Psi_A(\tau_{1})}\right]\label{eq:tcomplete2} \\
   \ln\left[\frac{\Psi_B(\tau)}{2q^2 e^{q\tau}}\frac{\Psi(\tau_{1})}{\Psi_A(\tau_1)}\frac{\Psi_A(\tau_3)}{\Psi_B(\tau_3)}\right].\label{eq:tcomplete3}
\end{subnumcases}

In the same manner we obtain an expression for $r(\tau)$:
\begin{subnumcases}{\label{eq:xcomplete}r(\tau) = }
			  2\alpha q\frac{H-Se^{q\tau}}{HS\Psi(\tau)}+r_{\infty} \label{eq:xcomplete1}\\
			  -2\alpha q\frac{\Psi_A(\tau_{1})}{\Psi(\tau_{1})}\frac{H-DA_{\tau_{1}}e^{q\tau}}{HDA_{\tau_{1}}\Psi_A(\tau)} \\ 
				  \nonumber\quad\quad + \mathcal{K}(\tau_1) + r_\infty \label{eq:xcomplete2}\\
			  2\alpha q\frac{\Psi_A(\tau_1)}{\Psi(\tau_1)}\frac{\Psi_B(\tau_3)}{\Psi_A(\tau_3)}\frac{H-DB_{\tau_3}e^{q\tau}}{HDB_{\tau_3}\Psi_B(\tau)} \\
				  \nonumber\quad\quad + \mathcal{H}(\tau_3)  + \mathcal{K}(\tau_1) + r_\infty \label{eq:xcomplete3}
\end{subnumcases}
with
\begin{equation}
\begin{aligned}
\mathcal{K}(\tau_{1}) &= \frac{2\alpha q}{\Psi(\tau_{1})}\frac{-2DSA_{\tau_1}e^{q\tau_1} + H[S+A_{\tau_1}D]}{HDSA_{\tau_1}} \\ 
\mathcal{H}(\tau_3)   &= \frac{2\alpha q}{\Psi(\tau_1)} \frac{\Psi_A(\tau_1)}{\Psi_A(\tau_3)}\frac{2DA_{\tau_1}B_{\tau_3}e^{q\tau_3}-H[A_{\tau_1} + B_{\tau_3}]}{HDA_{\tau_1}B_{\tau_3}}
\end{aligned}
\label{eq:KK}
\end{equation}
and 
\begin{equation}
 r_\infty = \frac{\alpha}{HS} = \frac{c}{H}\frac{1}{\sqrt{1+(Hc/\alpha)^2}+Hc/\alpha}.
\label{eq:xinfty}
\end{equation}

Here, $r_\infty$ is Rindler's ``$\alpha$-horizon``, which is the maximum coordinate distance that Tina can reach asymptotically when she accelerates away from Earth for all times starting at time $t_i=0$ with zero initial velocity. Note that $r_{\infty}<r_{\text{\tiny f1}}=c/H$, cf. Eq.~\ref{eq:critForwardPoints}.

\subsection{Analysis of the worldline}

\subsubsection{Comparison with flat space}
The worldline $(t_{\text{\tiny flat}}, r_{\text{\tiny flat}})$ in flat space (see Eqs.~(\ref{eq:tMink}) and (\ref{eq:xMink})), is a special case for $H\rightarrow 0$ 
of the more general worldline in de Sitter space. In order to show the similarity of the respective expressions, we rewrite Eqs.~(\ref{eq:tcomplete1}) 
and (\ref{eq:xcomplete1}) and obtain
\begin{equation}
 r(\tau) = \frac{c^2}{\alpha}\frac{e^{q\tau}+e^{-q\tau}-2}{2+\zeta(\tau)}, \quad t(\tau)= \frac{1}{H}\ln\left(\frac{2+\zeta(\tau)}{2q^2\frac{c^2}{\alpha^2}}\right),
\end{equation}
with
\begin{equation}
 \zeta(\tau) = H\frac{c^2}{\alpha^2}\left(Se^{q\tau} - De^{-q\tau}\right).
\end{equation}
Because $\zeta(\tau)=0$ and $q=\alpha/c$ for $H=0$, we can directly see that $r(\tau)\rightarrow r_{\text{\tiny flat}}(\tau)$ for $H\rightarrow 0$ and 
that $\zeta(\tau)$ characterizes the deviation of the $r(\tau)$-coordinate function from flat space for $H>0$, except for the difference of $q$ 
and $\alpha/c$.

The situation is more difficult for $t(\tau)$. Using L'Hopital's law we can evaluate $t(\tau)$ in the limit $H\rightarrow 0$ and, indeed, obtain $t(\tau)\rightarrow t_{\text{\tiny flat}}(\tau)$ in this case. The deviation from the flat space worldline is very small in the beginning, as $H$ is very small. The current value of the Hubble constant is (see Hinshaw et al.\cite{WMAP5})
\begin{equation}
 H_{0}\approx (70.5\pm1.3)\frac{\text{km}}{\text{s}\cdot\text{Mpc}} \approx (7.21\pm 0.13)\times 10^{-11}\frac{\text{ly}}{\text{y}\cdot\text{ly}},
\label{eq:H0}
\end{equation}
Thus the effect of the expansion is negligible for trips within our galaxy, for example (of course we do not live in a de Sitter universe after all).
However, the properties of the worldline for $\tau\rightarrow\infty$ change considerably. Most importantly, for $H>0$, $r(\tau)$ has the upper boundary 
$r(\tau\rightarrow\infty) = r_{\infty}$, as already mentioned.

\subsubsection{Round trip with equal acceleration and deceleration times}
In this section we consider a simple trip, where Tina chooses her acceleration and deceleration times equally long, i.e. $\tau_3 = 3\tau_1$. As in flat space we choose $\tau_1=5$~y. To illustrate the influence of the expansion during such a short trip, we choose an extremely large Hubble constant. Specifically, we consider the cases $H= \{10^{7} H_{0}, 10^{8} H_{0}, 10^{9} H_{0}\}$, where $H_{0}$ is the Hubble constant of our universe, cf. Eq.~(\ref{eq:H0}).

\begin{figure}[htb]
 \includegraphics[width=3.4in]{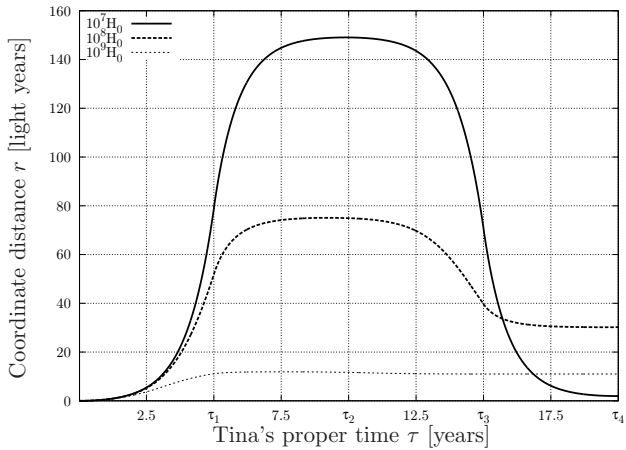}
 \caption{Coordinate distance $r$ for trips with $\tau_1=5$~y, $\tau_3=15$~y in de Sitter space for very large Hubble constants.}
 \label{fig:XSitter}
\end{figure}

\begin{figure}[htb]
 \includegraphics[width=3.4in]{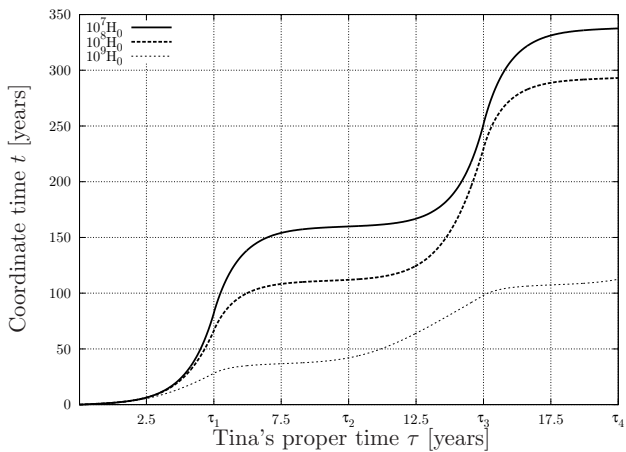}
 \caption{Coordinate time $t$ for the same trips as in Fig.~\ref{fig:XSitter}.}
 \label{fig:TSitter}
\end{figure}

\begin{figure}[htb]
 \includegraphics[width=3.4in]{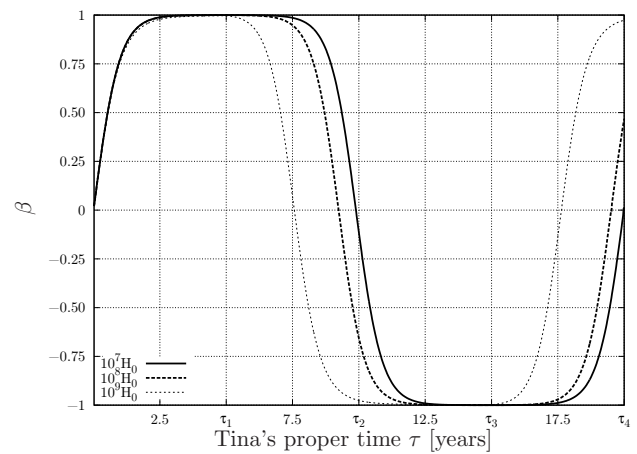}
 \caption{Velocity $\beta=v/c$ for the same trip as in Fig.~\ref{fig:XSitter}. Contrary to flat space $\beta\neq 0$ for $\tau=20$~y.}
 \label{fig:betaSitter}
\end{figure}

Figure~\ref{fig:XSitter} shows Tina's radial coordinate $r(\tau)$ during these journeys. Even for a Hubble constant that is $10^7$ times larger than that of our universe, the expansion has only minor influence on the course of Tina's journey, cf. Fig.~\ref{fig:XMink}. In that case, she reaches her maximum radial coordinate distance $r_{\text{\tiny}}\approx 149.06$~ly at proper time $\tau\approx 9.8898$~y. After $4\tau_1=20$~y, however, she has not returned yet, because $r\approx 1.9511$~ly.
It is not surprising, that $r(\tau_4)>0$. As the spacetime expands, Tina cannot return to Earth on these journeys. Besides that, there are several other differences to Minkowski space.
Figure~\ref{fig:TSitter} shows the elapsed coordinate time $t(\tau)$ during the same trips. The expansion rate has a strong influence on time dilatation. As we have shown in Eq.~(\ref{eq:maxgamma}), the maximum possible velocity $\beta_\infty$ and the maximum value $\gamma_\infty$ of $t'$ become smaller when $H$ increases. The same is also true for a round trip. 
Thus, time dilatation is the smaller, the larger the expansion rate is. In Table~\ref{fig:maxtable}, $r_\infty$ and $\gamma_\infty$ are compared for the expansion rates considered here.
In Fig.~\ref{fig:betaSitter}, Tina's velocity $\beta=v/c$ during her trip is shown. Clearly, $\beta=0$ already for some time $\tau<2\tau_1$ and $\beta\neq 0$ at the end of the trip.
In Table~\ref{fig:rettable}, the elapsed coordinate times $t(\tau_4)$ after these journeys and the respective journey in Minkowski space, as well as the coordinate distance $r(\tau_4)$, the proper distance $l(\tau_4)$, and the velocity $\beta(\tau_4)$ are compared. For $H=10^7 H_0$ the difference of
coordinate time is small, but for even larger Hubble constants the elapsed time in de Sitter space is considerably smaller. On the other hand, Tina's velocity at the end of the trip is not zero
and increases significantly with $H$, as well as her proper distance. Her coordinate distance is also not zero with a more complex dependence on $H$. This is due to the fact that $r_\infty$ becomes smaller for larger $H$ and therefore also 
the coordinate distances that Tina can reach during the respective journeys.

\begin{table}[ht]
\centering\setlength{\tabcolsep}{0.15cm}
\begin{tabular}{c | c c c }
$H$ & $r_\infty$[ly] & $\gamma_\infty$ & $\beta_\infty$ \\ [0.5ex]
\hline\\[-0.8em]
$H_0$     & $1.3869\times 10^{10}$& $1.4322\times 10^{10}$   & $1-2.4376\times 10^{-21}$  \\
$10^7H_0$ & $1.3860\times 10^{3}$ & $1.4322\times 10^{3}$ & $1-2.4376\times 10^{-7}$ \\
$10^8H_0$ & $1.3773\times 10^{2}$ & $1.4322\times 10^{2}$ & $1-2.4375\times 10^{-5}$ \\
$10^9H_0$ & $1.3773\times 10^{1}$ & $1.4322\times 10^{1}$ & $1-2.4287\times 10^{-3}$ \\
\end{tabular}
\caption{Maximum radial coordinate $r_\infty$, Lorentz-factor $\gamma_\infty$, and velocity $\beta_\infty$ for $\alpha=9.81$~$\text{m}/\text{s}^2\approx 1.0326$~$\text{ly}/\text{ly}^2$ after infinite acceleration time in universes with different 
Hubble constants.}
\label{fig:maxtable}
\end{table}

\begin{table}[ht]
\centering\setlength{\tabcolsep}{0.25cm}
\begin{tabular}{c | c c c c}
$H$ & $t(\tau_4)$[y] & $r(\tau_4)$[ly] & $l(\tau_4)$[ly] & $\beta(\tau_4)$  \\ [0.5ex]
\hline\\[-0.8em]
Minkowski     & 338.36 &  0 & 0  & 0\\
$10^7H_0$ & 337.55 & 1.95 & 2.49 & 0.0130\\
$10^8H_0$ & 293.09 & 30.21 & 249.98 & 0.4668\\
$10^9H_0$ & 112.45 & 10.99 & 36456.47 & 0.9736\\
\end{tabular}
\caption{Elapsed time $t(\tau_4)$, coordinate distance $r(\tau_4)$ and proper distance $l(\tau_4)$ and velocity $\beta(\tau_4)$ at the and of the trips in Fig.~\ref{fig:XSitter} and the equivalent trip in Minkowski space in Fig.~\ref{fig:XMink}.}
\label{fig:rettable}
\end{table}
One further aspect can be seen in Fig.~\ref{fig:TSitter}. From flat space, one would expect that $t(4\tau_1)=2t(2\tau_1)$. Thus, the elapsed coordinate time after stage~\ding{175}
is twice the time after stage~\ding{173}, but here this is different. 
\\
The reason for these discrepancies are the different times Tina needs to accelerate to reach a certain velocity, and the time Tina needs to decelerate to come to rest again.
 We will evaluate this further in the next section.

\subsubsection{Time to come to rest}\label{sec:TimeToRest}
In the preceding section we have shown that a journey with four stages of the same duration is not an appropriate choice in de Sitter space. To perform a proper round trip, we must find the duration of stage~\ding{173} necessary for a given stage~\ding{172} such that, at its end, Tina is at rest again. Then, stages~\ding{174} and \ding{175} have to be chosen in such a way that she returns to Earth and arrives there with zero velocity.

When Tina is at rest, we have $t'=1$. With Eq.~(\ref{eq:tscomplete2}) we can use this condition to calculate $\tau_{\text{rest}}$ and obtain
\begin{equation}
 \tau_{\text{rest}} = \frac{1}{q}\ln\dfrac{S}{A_{\tau_{1}}D}
\label{eq:taurest}
\end{equation}
for the duration of stages~\ding{172} + \ding{173}. It can easily be seen that 
\begin{equation}
 \tau_{\text{rest}}<2\tau_1.
\end{equation}
Thus, in de Sitter space stage~\ding{173} is always shorter than stage~\ding{172}.
\\
In the limit $\tau_1\rightarrow\infty$, we obtain
\begin{equation}
 \lim_{\tau_{1}\rightarrow\infty}(\tau_{\text{rest}}-\tau_{1}) = \frac{1}{q}\ln\frac{S}{H}
\label{eq:limtau2}
\end{equation}
for the duration of stage~\ding{173}.
Thus, the time needed to come to rest after accelerating for arbitrarily long times has an upper boundary!

\subsubsection{Maximum acceleration time\label{sec:MaxAccTime}}
If Tina accelerates too long in the beginning of her trip, she can no longer return home to Earth afterwards. 
The condition for a possible return is $l(\tau_{\text{rest}}) < r_\infty$ for Tina's proper distance at the end of stage~\ding{173}.

To make this point clear, we assume Tina is on a journey and has covered a proper distance $l(\tau_{\text{rest}}) = r_\infty$. 
A transformation to new coordinates
\begin{equation}
 t\mapsto t-t(\tau_{\text{rest}}), \quad r\mapsto [r-r(\tau_{\text{rest}})]e^{Ht(\tau_{\text{rest}})},
\end{equation}
puts the previous expansion of the universe into the definition of our new coordinates. This transformation makes sense because of the form of the expansion factor, $R(t)=e^{Ht}$. Hence, we have $R(t_b)/R(t_a) = R(t_b-t_a)$ for arbitrary $t_a$ and $t_b$, see also Tolman.\cite{Tolman1934} 
Thus, current proper distances in the old coordinates again equal coordinate distances in the new coordinates and the equations of motion are the same in the new coordinates.

In these new coordinates, Tina's current position is $r=0$ and Earth is located at $r_{\text{Earth}} = r_\infty$. Therefore, she can no longer return home. Hence the set of all particles at rest at different radial coordinates $r$ can be divided into four subsets at the beginning of the journey:
\begin{enumerate}
 \item Particles beyond Tina's future light cone.
 \item Particles, which Tina cannot reach because they have $r\geq r_\infty$.
 \item Particles, which Tina can reach but where she cannot return to Earth afterwards, because when she arrives there, her proper distance to Earth is larger than $r_\infty$.
 \item Particles, which Tina can reach with $l<r_\infty$ and where she therefore can return to Earth afterwards.
\end{enumerate} 
For times of departure $t_0\neq 0$ the same classification can be made by replacing coordinate distances via $r\rightarrow re^{-Ht_0}$.
\\
To find the maximum acceleration time $\tau_{1\text{max}}$ that allows Tina to return home, we calculate the acceleration time for which Tina has exactly covered a proper distance
\begin{equation}
 l(\tau_{\text{rest}}) = e^{Ht(\tau_{\text{rest}})}r(\tau_{\text{rest}}) = r_\infty,
\label{eq:tau1max}
\end{equation}
at the end of stage~\ding{173}. Note that Eq.~(\ref{eq:tau1max}) is not a conditional equation for $\tau_{\text{rest}}$ but for $\tau_{1\text{max}}$, with $\tau_{\text{rest}}=\tau_{\text{rest}}(\tau_{1\text{max}})$! Solving this equation yields
\begin{equation}
 \tau_{1\text{max}} = \frac{1}{q}\ln\left(\frac{S}{H}\right).
\label{eq:taumax}
\end{equation}
Round trips are only possible for $\tau_1<\tau_{1\text{max}}$. For longer acceleration times Tina cannot return to Earth.
Using Eq.~(\ref{eq:taumax}), we further obtain
\begin{equation}
 \tau_{\text{restmax}} = \tau_{\text{rest}}(\tau_{1\text{max}}) = \frac{1}{q}\ln\left(\frac{S}{D}\frac{Hq+\frac{\alpha^2}{c^2}-H^2}{2H^2}\right).
\label{eq:taurestmax}
\end{equation}
For $H=H_0$ we obtain $\tau_{1\text{max}}=22.6463$~y and $\tau_{\text{rest}}(\tau_{1\text{max}}) = 45.2773$~y.
For details on the calculation see App.~\ref{appsec:AccTime}.

\begin{figure}[htb]
 \includegraphics[width=3.4in]{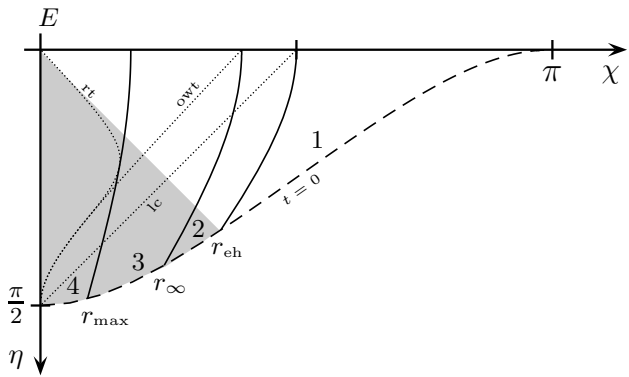}
 \caption{Situation in a de Sitter universe with $H=5\times 10^{9}H_0$. The solid lines represent the worldlines of particles at $r_{\text{max}}$, $r_\infty$, and $r_{\text{eh}}$, which separate the $t=0$ hypersurface into four regions. The dotted lines represent Tina's worldline on a round trip (rt) with $\tau_1=(1-10^{-8})\tau_1$, a one-way trip (owt) where she accelerates away from Earth for all times, and her future light cone (lc) at time $t=0$. 
The worldlines of all particles in region 1 $(r\geq r_{\text{eh}})$ do not intersect Tina's future light cone and Eric's backward light cone (gray-shaded region). Region 2 $(r_{\infty}\leq r<r_{\text{eh}})$ are all particles where Tina can send a light signal to but which cannot be reached by her. All particles in region 3 $(r_{\text{max}}\leq r<r_{\infty})$ can be reached by Tina, but she cannot return to Earth. Finally, all particles with $r<r_{\text{max}}$ (region 4), are possible destinations for a round trip starting at $t=0$.}
\label{fig:confRegions}
\end{figure}

This situation can easily be illustrated using CE coordinates. To receive Tina's worldline $[\eta(\tau), \chi(\tau)]$ for infinitely long acceleration and on a round trip, we insert Eqs.~(\ref{eq:tcomplete1}), (\ref{eq:xcomplete1}) and (\ref{eq:tcomplete}), (\ref{eq:xcomplete}),
respectively, into Eq.~(\ref{eq:coordTransf2}). Details on the calculation of $\tau_3$ and $\tau_4$ for a suitable round trip are given in the next section. Figure~\ref{fig:confRegions} shows Tina's worldline on a round trip (rt) with $\tau_1=(1-10^{-8})\tau_{1\text{max}}$, her worldline on a one-way trip (owt) where she accelerates away from Earth for all times, and her future light cone at time $t=0$ (lc) for a universe with $H=5\times 10^9 H_0$. In addition, the worldlines of particles at rest at Tina's event horizon $r_{\text{eh}}=c/H$ for $t=0$, cf. Eq.~(\ref{eq:EventHorizon}), at the $\alpha$-horizon $r_\infty$ and at $r_{\text{max}} \equiv r(\tau_{\text{restmax}})$ are depicted. Here, the particle at $r_{\text{max}}$ has the smallest distance, that Tina
cannot reach on a round trip. 

With Eqs.~(\ref{eq:tau1max}) and (\ref{eq:taurestmax}) we arrive at
\begin{equation}
 r_{\text{max}} = r_\infty e^{-Ht(\tau_{\text{restmax}})}.
\end{equation}
For $H=5\times 10^9 H_0$ we obtain $r_{\text{max}} = 0.40067r_\infty$, for the general expression see App.~\ref{appsec:Rmax}.
Tina's future light cone intersects the worldline of the particle at the event horizon (eh) at $\eta=0, \chi_{\text{eh}}=\pi/2$. Thus, this is the boundary particle of region 2 of particles Tina can send light signals to at time $t=0$. Equally, the worldline for a trip with infinitely long acceleration (owt) intersects the worldline of the particle at $r_\infty$ at $\eta=0, \chi_{\infty}=\arctan\left[\alpha/(Hc)\right]$, which borders region 3. On a round trip (rt) with $\tau_1$ very close to $\tau_{1\text{max}}$, Tina almost reaches the particle at $r_{\text{max}}$, which is the boundary particle of region 4 of possible destinations for round trips starting at $t=0$.

\subsubsection{A suitable round trip}
In this section, we study how long Tina has to choose stages~\ding{174} and \ding{175}, to reach Earth again and come to rest there at the end of her journey. The only difference between the outward trip and the return trip is the larger expansion factor $R[t(\tau_{\text{rest}})]$ instead of $R(0)=1$.
We use the same transformation as in the preceding section and consider Tina's proper distance $l(\tau_{\text{rest}})$ as a coordinate distance for $\tau=t=0$. Then we can, in principle, calculate the proper duration $\tau_{1\text{return}}$ for stage \ding{174}, which she needs to return home by calculating the time needed to cover the respective coordinate distance starting at $t=0$. The equation
\begin{equation}
 r[\tau_{\text{rest}}(\tau_{1\text{return}})] = d
\end{equation}
with arbitrary $d<r_\infty$ is more difficult to solve than the similar Eq.~(\ref{eq:tau1max}) for $l$, where the factor $e^{Ht}$ and the restriction on $d=r_\infty$ nicely simplify the occurring expressions. Therefore, this equation can only be treated numerically.
\\
The effects of expansion become especially clear for journeys with acceleration times close to $\tau_{1\text{max}}$.

As an example, we consider a universe with $H=10^9H_0$. From Eq.~(\ref{eq:taumax}) we obtain ${\tau_{1\text{max}}=2.597\,\text{y}}$ in this case. For our discussion, we consider three journeys, with acceleration times $\tau_1 = \{0.99\tau_{1\text{max}}, 0.9999\tau_{1\text{max}}, 0.99999\tau_{1\text{max}}\}$ 
in stage~\ding{172}. In Table \ref{fig:timetable}, these journeys are compared with respect to Tina's and Eric's elapsed times and Tina's maximum coordinate and proper distances at the end of stage~\ding{173}.
As the duration of stage~\ding{172} is chosen very close to the maximum acceleration time,
the time needed to return varies strongly with minimally different durations for stage~\ding{172}.

\begin{figure}[htb]
 \includegraphics[width=3.4in]{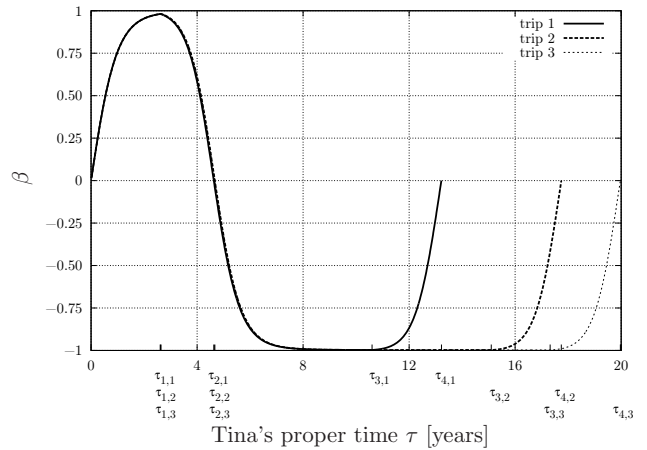}
 \caption{Velocity $\beta(\tau)$ for journeys with acceleration times $\tau_1 = 0.99\tau_{1\text{max}}$ (trip 1), $0.9999\tau_{1\text{max}}$ (trip 2), and $0.99999\tau_{1\text{max}}$ (trip 3). For comparison, the respective times $\tau_{n,m}$ for the end of phase $n$ in trip $m$ are marked.}
 \label{fig:BetaSitterprop}
\end{figure}

\begin{figure}[htb]
 \includegraphics[width=3.4in]{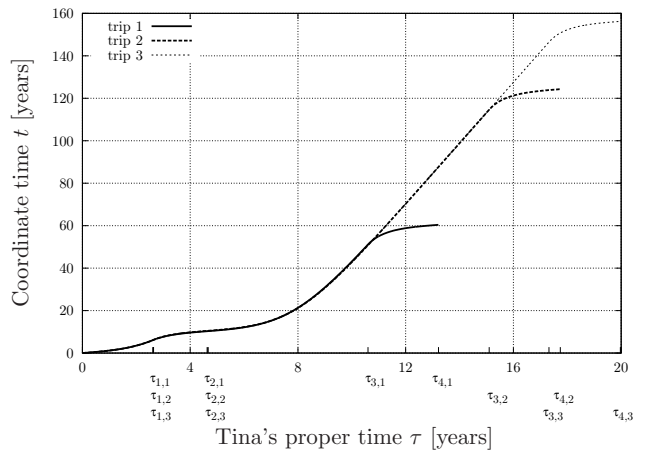}
 \caption{Coordinate time  $t$ for the same journeys as in Fig.~\ref{fig:BetaSitterprop}.}
 \label{fig:TSitterprop}
\end{figure}

\begin{figure}[htb]
 \includegraphics[width=3.4in]{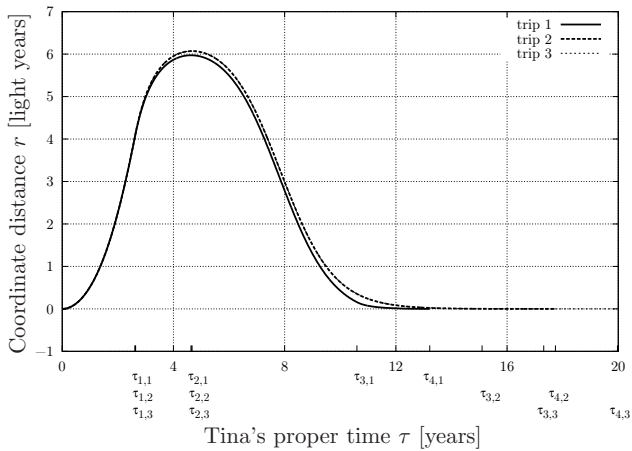}
 \caption{Coordinate distance $r$ for the same journeys as in Fig.~\ref{fig:BetaSitterprop}.}
 \label{fig:XSitterprop}
\end{figure}

\begin{figure}[htb]
 \includegraphics[width=3.4in]{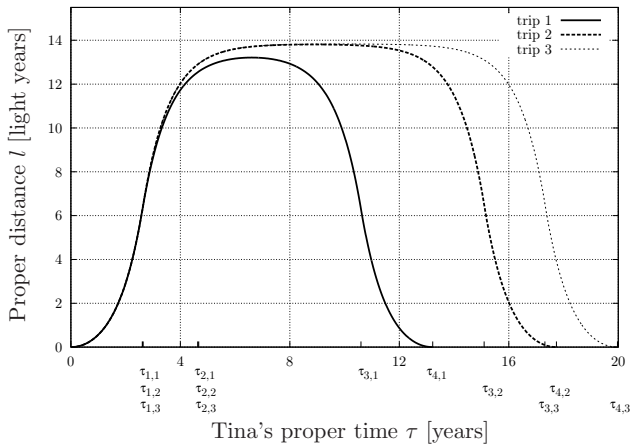}
 \caption{Proper distance $l$ for the same journeys as in Fig.~\ref{fig:BetaSitterprop}.}
 \label{fig:LSitterprop}
\end{figure}

Figures~\ref{fig:BetaSitterprop}--\ref{fig:LSitterprop} show $\beta(\tau)$, $t(\tau)$, $r(\tau)$, and $l(\tau)$ for these trips.
\begin{table*}[ht]
\centering\setlength{\tabcolsep}{0.25cm}
\begin{tabular}{c | c c c c | c c c | c c}
$\tau_1/\tau_{1\text{max}}$ & $\tau_2$[y] & $\tau_4-\tau_2$ [y] & $\tau_3$ [y] & $\tau_4$ [y] & $t(\tau_2)$ [y] &  $t(\tau_4)-t(\tau_2)$ [y] &  $t(\tau_4)$ [y] & $r(\tau_2)$ [$r_\infty$] & $l(\tau_2)$ [$r_\infty$] \\ [0.5ex]
\hline\\[-0.8em]
$0.99000$ & 4.6302 & 8.5862  & 10.60391 & 13.2164 & 10.2857 & 50.1219  & 60.4076  & 0.46173 & 0.96930 \\
$0.99990$ & 4.6686 & 13.0742 & 15.10422 & 17.7428 & 10.4866 & 113.8086 & 124.2952 & 0.46935 & 0.99969 \\
$0.99999$ & 4.6690 & 15.2990 & 17.32912 & 19.9679 & 10.4885 & 145.7429 & 156.2314 & 0.46942 & 0.99997 \\
\end{tabular}
\caption{Journeys with initial acceleration time $\tau_1$ close to the maximum acceleration time $\tau_{1\text{max}}$ for $H=10^{9}H_0$. The Table shows Tina's elapsed time $\tau_2$ and Eric's elapsed time $t(\tau_2)$ for the outward journey, Tina's time $\tau_3$ at the end of stage \ding{174}, Tina's time $(\tau_4-\tau_2)$ and Eric's time $t(\tau_4)-t(\tau_2)$ for the return journey and for the round trip $\tau_4$, $t(\tau_4)$, as well as the radial and proper distance $r(\tau_2)$, $l(\tau_2)$, which Tina 
covers during these journeys, compared to $r_\infty$.}
\label{fig:timetable}
\end{table*}


Tina has to travel with very high velocity for a very long time on her way home, thus causing a large time dilatation, see Fig.~\ref{fig:BetaSitterprop}. 
Therefore, the elapsed coordinate time after the trip also increases strongly with minimal increase in initial acceleration time, see Fig.~\ref{fig:TSitterprop}.
The comparison of Figs.~\ref{fig:XSitterprop} and \ref{fig:LSitterprop} shows the minor meaning of the r-coordinate as a measure of distance. For most of the return trip Tina's radial coordinate distance to Earth is very small. Her proper distance however has almost the same qualitative $\tau$-dependence as the coordinate distance in flat space, cf. Fig.~\ref{fig:XMink}.

\subsubsection{A trip to the end of the universe}
In flat Minkowski space, as already discussed by M\"uller et al\cite{Mueller2006}, Tina could reach the most distant galaxies about $1.37\times 10^{10}$~ly away on a trip with $\tau_1 \approx 22.635$~y. This can easily be reproduced by setting $\tau = 2\tau_1$ in Eq.~(\ref{eq:xMink2}). When Tina has reached her destination, however, $t\approx 1.37\times 10^{10}$~y have elapsed for Eric.

In a de Sitter universe the expansion has an extremely large effect during such a long trip, as is easily seen by evaluating
\begin{equation}
 r_\infty = 1.38696\times 10^{10}\,\text{ly},
\end{equation}
for $H=H_0$. Thus, the galaxies considered in flat space are out of reach for Tina, in fact they are beyond the event horizon, which differs only minimally from $r_\infty$ in this case, as $\alpha/c\ggg H_0$, cf. the last paragraph in Sec.~\ref{sec:WLcalc}.
Therefore, we consider the quasar\cite{3C324} 3C~324 with a distance of approximately $1.21\times 10^{10}$~ly (This is the distance a light ray emitted today would travel to the quasar if the expansion of the universe stopped today).
As less distant destinations we use the Andromeda Galaxy, which is approximately $2.56\times 10^{6}$~ly away, see e.g. McConnachie,\cite{McConnachie2005} the quasar\cite{Uchiyama2006} 3C~273 with a distance of approx. $2.44\times 10^{9}$~ly and the quasar\cite{3C147} 3C 147, which is approx. $6.44\times 10^{9}$~ly away.
 
Table~\ref{fig:flattable} shows how long Tina has to choose stage~\ding{172} to reach these destinations in Minkowski space. To make the comparison with de Sitter space easier, we
also list the times $\tau_2=2\tau_1$, $\tau_3=3\tau_1$, and $\tau_4=4\tau_1$ at the end of the respective stages. We also list how much time has elapsed for Eric when Tina has reached her destination, and when she has returned to Earth.
\begin{table*}[t]
\centering
\setlength{\tabcolsep}{0.25cm}
\begin{tabular}{c | c c c c c c c}
 Destination & Distance [ly] & $\tau_1$ [y] & $\tau_2$ [y] & $\tau_3$ [y] & $\tau_4$ [y] & $t(\tau_2)$ [y] & $t(\tau_4)$ [y] \\ [0.5ex]
\hline\\[-0.8em]
Andromeda & $2.56\times 10^{6}$  & 14.32048 &  28.64097 & 42.96145 & 57.28193 & $2.56000\times 10^{6}$ &  $5.12000\times 10^{6}$ \\
3C 273 & $2.44\times 10^{9}$  & 20.96353 & 41.92706 & 62.89060 & 83.85413 & $2.44000\times 10^{9}$ &  $4.88000\times 10^{9}$ \\
3C 147 & $6.44\times 10^{9}$  & 21.90340 & 43.80681 & 65.71021 & 87.61362 & $6.44000\times 10^{9}$ &  $1.28800\times 10^{10}$ \\
3C 324 & $1.21\times 10^{10}$ & 22.51183 & 45.02367 & 67.54248 & 90.04734 & $1.20710\times 10^{10}$ & $2.41420\times 10^{10}$ \\
\end{tabular}
\caption{Round trips in flat space. The Table shows Tina's acceleration times $\tau_1$ necessary to travel certain distances, her travel time $\tau_2$ to the destination, the time $\tau_3$ when she starts to decelerate on her way back to Earth and $\tau_4$ for the round trip, as well as Eric's elapsed time $t(\tau_2)$ when she reaches her destination, and $t(\tau_4)$ when she is back at Earth.}
\label{fig:flattable}
\begin{tabular}{c |  c c c c c c c }
 Destination & $\tau_1$ [y] & $\tau_2$ [y] & $\tau_3$ [y] & $\tau_4$ [y] & $t(\tau_2)$ [y] & $t(\tau_4)$ [y] & $l(\tau_2)$ $[r_\infty]$\\ [0.5ex]
\hline\\[-0.8em]
Andromeda & 14.32066 & 28.64115 & 42.96199  & 57.28265 & $2.56024\times 10^{6}$ & $5.12095\times 10^{6}$ & 0.00018\\
3C 273 & 21.15091 & 42.11445 & 63.49791 &  84.64882 &  $2.68367\times 10^{9}$ &  $6.01429\times 10^{9}$ & 0.21348\\
3C 147 & 22.50791 & 44.41132 & 68.87145 &  91.37936 &  $8.65778\times 10^{9}$ &  $3.66176\times 10^{10}$ & 0.86680\\
3C 324 & 24.48999 & 47.00182 & - & -  & $2.83312\times 10^{10}$ & -  & 6.71124 \\
\end{tabular}
\caption{The same numbers as in Table~\ref{fig:flattable} for round trips in de Sitter space with $H=H_0$. In addition, Tina's proper distance to Earth at her destination is listed.}
\label{fig:deSittertable}
\end{table*}
Table~\ref{fig:deSittertable} shows the same numbers in de Sitter space with $H=H_0$. Because of the expansion, stage~\ding{172} is longer than in flat space. Stage~\ding{173}, however,
is almost equally long as in flat space as Tina can decelerate more quickly than she accelerates, see Sec.~\ref{sec:TimeToRest}. The same is also true for the return journey, even to some greater extent.

For a trip to the Andromeda galaxy and even to the quasar 3C~273, the deviation from the flat space journey is very small. At these destinations, Tina's proper distance to 
Earth is still well less than $r_\infty$. For the quasar 3C~147, the deviations are very large, especially when comparing Eric's elapsed time in both spacetimes. This quasar
is almost at the maximum distance that allows Tina to return to Earth, when she arrives there, her proper distance to Earth is $l\approx 0.86680 r_\infty$.

For 3C~324, only a one way trip is possible, Tina cannot return to Earth afterwards. 
Also, Tina has to choose stage~\ding{172} more than two years longer than in flat space. As she moves at the highest velocity during these two years, this
rather small difference in acceleration time causes a huge difference on the trip. If Tina accelerates for the same time $\tau_1 = 24.48999$~y in Minkowski space, she covers a distance of $r = 9.30827\times 10^{10}$~ly at the end of stage~\ding{173}, which is almost eight times the distance with $\tau_1 = 22.51183$~y.

The huge influence of these additional two years can also be seen by comparing Eric's elapsed times in Minkowski space and de Sitter space. When Tina reaches 3C~324, more than twice the amount of time has elapsed in de Sitter space.

\begin{figure}[htb]
 \includegraphics[width=3.4in]{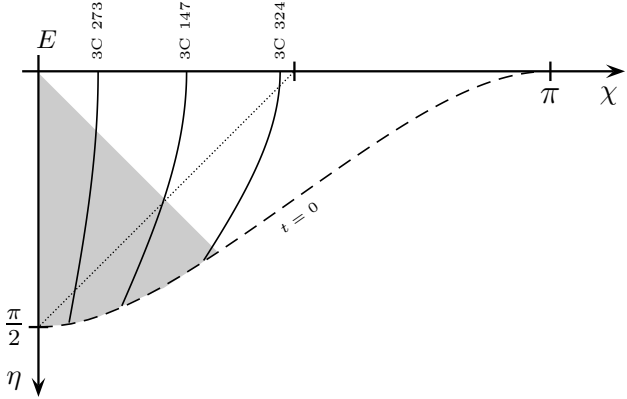}
 \caption{Worldlines of the quasars 3C~273, 3C~147, and 3C~324 in CE coordinates, $H=H_0$. The dotted line indicates Tina's forward light cone at $t=0$. The gray-shaded region corresponds to Eric's backward light cone. The quasar 3C~324 is close to Tina's future light cone.}
 \label{fig:confGalaxies}
\end{figure}

\section{Communication between the twins}\label{sec:Communication}
We imagine that Eric and Tina permanently send each other information about their respective current time $\tau$ or $t$.
We study this situation both from Tina's and Eric's perspective. Concretely, we investigate when a signal, sent by Eric at time $t_{\text{S}}$, will be received by Tina, and when a signal that Eric receives at time $t_{\text{R}}$ was sent by Tina. For Tina, we calculate the respective times $\tau_{\text{S}}$ and $\tau_{\text{R}}$. Again, we compare our results with Minkowski space.

\subsection{Infinitely long acceleration}
First we consider a journey where Tina accelerates away from Earth for all times. In Minkowski space a light signal sent by Eric at time $t_{\text{S}}$ is described by 
\begin{equation}
 r_{\text{L}}(t) = c(t-t_{\text{S}}).
\end{equation}
This light signal reaches Tina, when 
\begin{equation}
 r_{\text{L}}[t(\tau)] = r(\tau).
\end{equation}
With the expressions in Eqs.~(\ref{eq:tMink1}) and (\ref{eq:xMink1}) we obtain
\begin{equation}
 t_{\text{S}} = \frac{c}{\alpha}\left(1-e^{-\frac{\alpha}{c}\tau_{\text{R}}}\right), \quad\tau_{\text{R}} = -\frac{c}{\alpha}\ln\left(1-\frac{\alpha}{c}t_{\text{S}}\right).
\label{eq:tSendMink}
\end{equation}
Obviously Tina can only receive messages that Eric sends at times $t_{\text{S}}<c/\alpha$.

Light signals sent by Tina at her proper time $\tau_{\text{S}}$ are described by
\begin{equation}
\begin{aligned}
 r_{\text{L}}(t) &= r_{\text{flat}}(\tau_{\text{S}}) + ct_{\text{flat}}(\tau_{\text{S}}) - ct \\
	      &= \frac{c^2}{\alpha}\left(e^{\frac{\alpha}{c}\tau_{\text{S}}}-1\right) - ct,
\end{aligned}
\label{eq:Lightflat}
\end{equation}
where we have used Eqs.~(\ref{eq:tMink1}) and (\ref{eq:xMink1}). In this case we obtain
\begin{equation}
 \tau_{\text{S}} = \frac{c}{\alpha}\ln\left(1+\frac{\alpha}{c}t_{\text{R}}\right), \quad t_{\text{R}} = \frac{c}{\alpha}\left(e^{\frac{\alpha}{c}\tau_{\text{S}}}-1\right).
\label{eq:tauSendMink}
\end{equation}
Hence, Eric can receive every signal from Tina.

For radially moving light signals in de Sitter space, $(\vartheta,\varphi=\text{const})$, we have with Eq.~(\ref{eq:metric})
\begin{equation}
 dr = \pm ce^{-Ht}dt.
\label{eq:dlight}
\end{equation}
Light signals sent by Eric are moving outward $(dr>0)$, signals sent by Tina have $dr<0$. Thus, light signals sent by Eric at time $t_{\text{S}}$ are described via 
\begin{equation}
 r_{\text{L}}(t) = \frac{c}{H}\left(e^{-Ht_{\text{S}}} - e^{-Ht}\right),
\label{eq:xlight}
\end{equation}
where the constant of integration is chosen such that $r(t_\text{S})=0$.
\\
Tina receives these signals, when
\begin{equation}
 r_{\text{L}}[t(\tau_{\text{R}})] = r(\tau_{\text{R}}).
\label{eq:EricSignal}
\end{equation}
Inserting Eq.~(\ref{eq:tcomplete1}) into Eq.~(\ref{eq:xlight}) yields
\begin{equation}
 r_{\text{L}}(\tau_{\text{R}}) = \frac{c}{H}\left(e^{-Ht_{\text{S}}}-\frac{2q^2e^{q\tau}}{\Psi(\tau_{\text{R}})}\right).
\label{eq:xLighttau}
\end{equation}
Rearranging Eq.~(\ref{eq:EricSignal}) then leads to
\begin{subequations}
\begin{align}
 t_{\text{S}} &= -\frac{1}{H}\ln\left(2q\frac{\left(q-\frac{\alpha}{c}\right)e^{q\tau_{\text{R}}}+\frac{\alpha}{c}\frac{H}{S}}{\Psi(\tau_{\text{R}})}+\frac{\alpha}{cS}\right) \\
\nonumber &= \frac{1}{H}\ln\left(\frac{HSe^{q\tau_{\text{R}}} - \frac{\alpha}{c}\left(q-\frac{\alpha}{c}\right)}{\frac{\alpha}{c}He^{q\tau_{\text{R}}} + S\left(q-\frac{\alpha}{c}\right)}\right), \\
\tau_{\text{R}} &=\frac{1}{q}\ln\left(\frac{q-\frac{\alpha}{c}}{H}\frac{S+\frac{\alpha}{c}e^{-Ht_{\text{S}}}}{Se^{-Ht_{\text{S}}}-\frac{\alpha}{c}}\right)
\end{align}\label{eq:tSendSitter}\end{subequations}
as generalization of Eq.~(\ref{eq:tSendMink}).
\\
With $\Psi(\tau)$ from Eq.~(\ref{eq:RindlerPsi}) we obtain
\begin{equation}
 t_{\text{Smax}} = \lim_{\tau\rightarrow\infty}t_{\text{S}} = -\frac{1}{H}\ln\left(\frac{H}{c}r_{\infty}\right) = \frac{1}{H}\ln\left(\frac{c}{\alpha}S\right).
\label{eq:tSMax}
\end{equation}
Only signals sent by Eric at times $t<t_{\text{Smax}}$ can reach Tina. For $H\rightarrow 0$ we further obtain
\begin{equation}
\lim_{H\rightarrow 0}t_{\text{Smax}} = \frac{c}{\alpha},
\end{equation}
in accordance with the result for flat space. 

The dependence of $t_{\text{S}}$ on $H$ is very weak. For a universe with $H=H_0$ we find a deviation for $t_{\text{Smax}}$ around $1:10^{-21}$ from the flat 
space result, for $H=10^9 H_0$ it is still only a difference of $1:8.1\times 10^{-4}$. This is not surprising as we are on a very small timescale compared to times where the expansion has a measurable effect.

In Minkowski space every signal sent by Tina eventually reaches Eric. In de Sitter space this is not true, since Tina eventually crosses Eric's event horizon.
Tina's signals in de Sitter space are described via
\begin{equation}
 r_{\text{L}}(t) = \frac{c}{H}\left(e^{-Ht} - e^{-Ht(\tau_{\text{S}})}\right) + r(\tau_{\text{S}}), 
\end{equation}
instead of Eq.~(\ref{eq:Lightflat}) and by the expressions in Eqs.~(\ref{eq:tcomplete1}), (\ref{eq:xcomplete1}). We obtain
\begin{subequations}
\begin{align}
\tau_{\text{S}} &=\frac{1}{q}\ln\left(\frac{q+\frac{\alpha}{c}}{H}\frac{S-\frac{\alpha}{c}e^{-Ht_{\text{R}}}}{Se^{-Ht_{\text{R}}}+\frac{\alpha}{c}}\right), \\
 t_{\text{R}} &= -\frac{1}{H}\ln\left(2q\frac{\left(q+\frac{\alpha}{c}\right)e^{q\tau_{\text{S}}}-\frac{\alpha}{c}\frac{H}{S}}{\Psi(\tau_{\text{S}})} - \frac{\alpha}{cS}\right) \\
\nonumber &=\frac{1}{H}\ln\left(\frac{HSe^{q\tau_{\text{S}}}+\frac{\alpha}{c}\left(q+\frac{\alpha}{c}\right)}{S\left(q+\frac{\alpha}{c}\right) - \frac{\alpha}{c}He^{q\tau_{\text{S}}}}\right)
\end{align}\label{eq:tauSendSitter}\end{subequations}
instead of Eq.~(\ref{eq:tauSendMink}).
\\
When Tina crosses Eric's event horizon, $t_{\text{R}}$ diverges. Hence, we can use it to calculate the respective proper time by setting the argument of the logarithm equal to zero.
The result of this calculation is
\begin{equation}
 \tau_{1\text{eh}} = \frac{1}{q}\text{acosh}\left(1+\frac{c}{\alpha}\frac{q^2}{H}\right) = \frac{1}{q}\ln\left[\frac{S}{H}\left(q\frac{c}{\alpha}+1\right)\right].
\end{equation}
For $H=H_0$ Tina has to accelerate $23.3176$~y to cross Eric's event horizon, for $H=10^9 H_0$ it takes her $3.3096$~y.

\begin{figure}[htb]
 \includegraphics[width=3.4in]{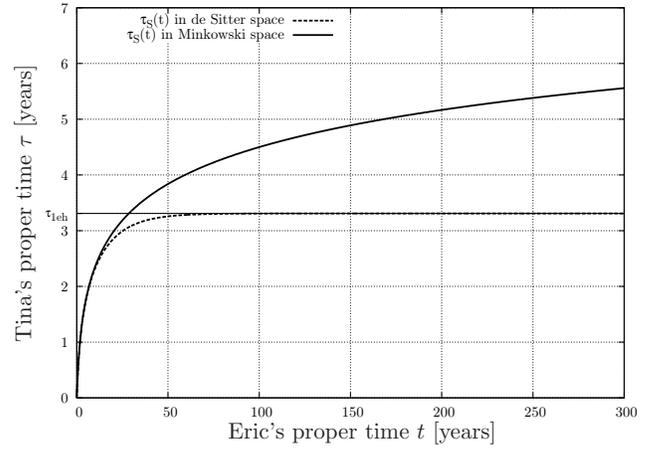}
 \caption{Eric's perspective: In Minkowski space, Eric has to wait very long for messages from Tina, but eventually he receives all signals from her. In de Sitter space $(H=10^9 H_0)$, Eric cannot receive signals that Tina
has sent after $\tau_{1\text{eh}}$. For all times, he therefore receives messages that Tina has sent at some earlier time.}
 \label{fig:TauSend}
\end{figure}

\begin{figure}[htb]
 \includegraphics[width=3.4in]{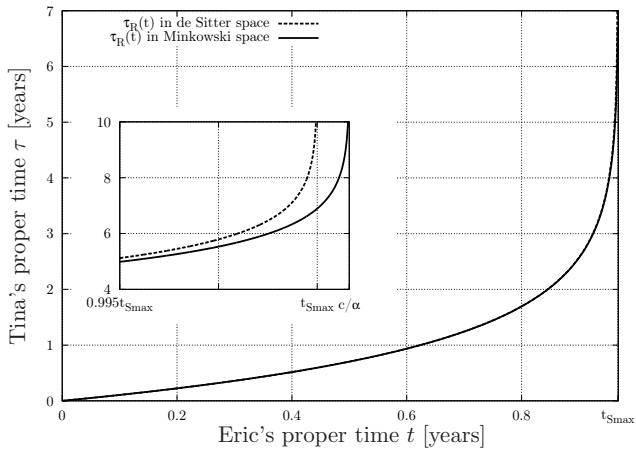}
 \caption{Eric's perspective: As Tina accelerates away from Eric, only messages that he sends before $t=c/\alpha$ in Minkowski space and before $t=t_{\text{Smax}}$ from Eq.~(\ref{eq:tSMax}) in de Sitter space $(H=10^9 H_0)$ can reach Tina. To show the effect of the expansion the region around $t_{\text{Smax}}$ is also shown enlarged. Only signals sent at times $t\lesssim t_{\text{Smax}}$ are received significantly later in de Sitter space. For $t\in[t_{\text{Smax}}, c/\alpha)$ communication is only possible in Minkowski space.}
 \label{fig:TauRec}
\end{figure}

In Figs.~(\ref{fig:TauSend}) and (\ref{fig:TauRec}) we look at Eric's perspective in a Minkowski space and a de Sitter space with $H=10^9H_0$. For signals sent by Eric, the effect of the expansion is hard to recognize. For signals which he receives from Tina,
the expansion has a large effect, when Tina approaches his event horizon and eventually moves behind it.

\subsection{Communication during a round trip}
To study communication during a round trip, the complete worldline of Tina has to be considered. The results are analogous to those in the preceding section.
We present figures to illustrate this situation and omit the explicit mathematical expressions.
For flat space, this problem was already considered by M\"uller\cite{Mueller2006} especially for a flight to Vega. 

Again we consider a trip with stage~\ding{172} of $\tau_1 = 0.99999\tau_{\text{max}}$ in a universe with $H=10^9 H_0$ and a trip in a flat universe with equally long stages~\ding{172}--\ding{175} for comparison.

\begin{figure}[htb]
 \includegraphics[width=3.4in]{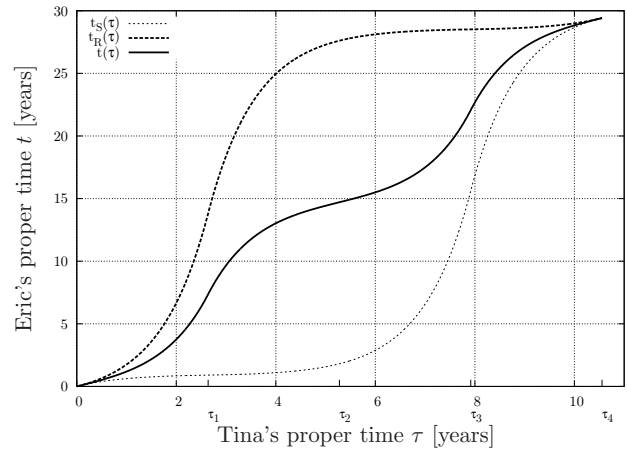}
 \caption{Tina's perspective in Minkowski space: At the proper time $\tau$ of Tina, the coordinate time is $t=t(\tau)$. A signal Tina receives at this time was emitted by Eric at time $t_{\text{S}}(\tau)$, and 
  a signal she emits at that time will be received by Eric at time $t_{\text{R}}(\tau)$.}
 \label{fig:AllTimesH0}
\end{figure}

\begin{figure}[htb]
 \includegraphics[width=3.4in]{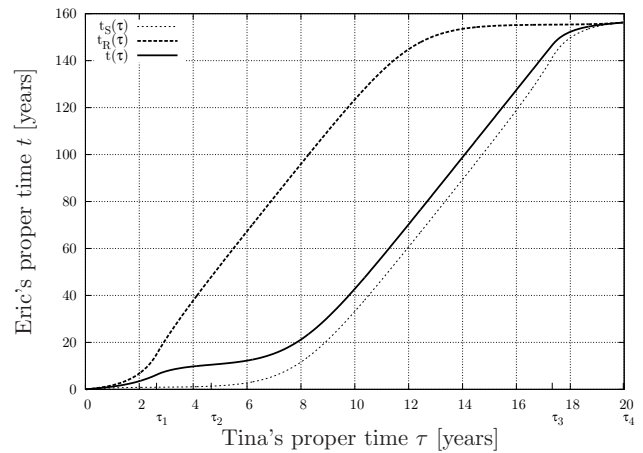}
 \caption{Same situation as in Fig.~\ref{fig:AllTimesH0} but in de Sitter space with $H=10^9 H_0$.}
 \label{fig:AllTimes0p99999}
\end{figure}

This time we consider the situation from Tina's perspective. Figure~\ref{fig:AllTimesH0} shows the situation in flat space, Fig.~\ref{fig:AllTimes0p99999} the situation in de Sitter space. In both cases, Tina receives very few signals
at the beginning of her journey. This is easy to understand, as she accelerates away from Earth, the travel time of Eric's signals increases rapidly and time dilatation additionally increases this effect. 
When she decelerates and starts to return to Earth, the rate of received signals increases. On the other hand most of her signals reach Eric only shortly before she herself returns to Earth.
The difference between the journeys in Minkowski space and de Sitter space is the large period in de Sitter space, where the rate of received signals remains constant for Tina and also for Eric. Also,
in flat space the travel time of the incoming lightray equals the travel time of the outgoing lightray, as they have to cover the same distance. Thus, in Fig.~\ref{fig:AllTimesH0} we always have 
\begin{equation}
t_{\text{R}}(\tau) - t(\tau) = t(\tau) - t_{\text{S}}(\tau).
\end{equation}
For $H=10^9 H_0$, however, the deviations are very large because Tina's signal has to travel a larger proper distance to reach Eric.

\section{Summary}
In this work we have studied the extension of the twin paradox to de Sitter space. We showed that an expanding space time has a huge influence on long journeys, and we were able
to quantitatively compare journeys in this spacetime with their counterparts in flat space concerning the duration of the respective journeys, the possibility of communication during the journeys,
and the limitations that exist for round trips due to the expanding spacetime, which can make a return to the point of departure impossible.

Because of its simple structure the de Sitter spacetime allows for extensive analytical calculations. The generalization of the investigations presented in this paper to more realistic spacetimes would be highly 
desirable. Since for such spacetimes not even the worldline can be determined analytically, all calculations would have to be performed numerically. Clearly, the results obtained for the twin paradox in de Sitter spacetime 
can serve as a guide in such calculations.  
 -----------------------------------------------------------------
\appendix

\section{Hyperbolic motion in Minkowski space}\label{appsec:hypmotionflat}
In flat Minkowski space, the four-acceleration $a^{\mu}$, see Eq.~(\ref{eq:amu}), for radial motion $(\vartheta, \varphi =\text{const})$ simplifies to
\begin{equation}
 a^{\mu} = \frac{d^{2}x^{\mu}}{d\tau^{2}}.
\end{equation}
Therefore, Eq. (\ref{eq:hypmotion}) yields
\begin{equation}
 \alpha^2 = -c^2t'' + r''^2.
\label{eq:hypmotionflat}
\end{equation}
On the other hand, the constraint $g_{\mu\nu}u^{\mu}u^{\nu} = -c^2$ yields
\begin{equation}
 r'^2 = c^2\left(t'^2-1\right).
\label{eq:constraintflat1}
\end{equation}
Differentiating both sides logarithmically and multiplying with $r'$ leads to
\begin{equation}
 r'' = \frac{t't''}{t'^2-1}r'.
\label{eq:constraintflat2}
\end{equation}
Combining Eqs.~(\ref{eq:hypmotionflat}), (\ref{eq:constraintflat1}), and (\ref{eq:constraintflat2}) yields the differential equation
\begin{equation}
t''^2 = \left(\frac{\alpha}{c}\right)^2\left(t'^2-1\right),
\label{eq:DiffEqFlat}
\end{equation}
with the solution $t'(\tau) = \cosh\left(\frac{\alpha}{c}\tau + c_{0} \right)$. Thus, with $t(0)=0$ and $r(0)=0$ we obtain
\begin{equation}
\textstyle t(\tau) = \frac{c}{\alpha}\sinh\left(\frac{\alpha}{c}\tau\right),\quad r(\tau) = \frac{c^2}{\alpha}\left[\cosh\left(\frac{\alpha}{c}\tau\right)-1\right],
\end{equation}
see Eqs.~(\ref{eq:tMink1}) and (\ref{eq:xMink1}). The derivation of Eqs.~(\ref{eq:tMink2}), (\ref{eq:tMink3}) and (\ref{eq:xMink2}), (\ref{eq:xMink3}) is straightforward.

\section{Rindler's Calculations}\label{appsec:preliminary}
With the relevant Christoffel symbols\cite{Mueller2009}
\begin{equation}
\Gamma_{tr}^{r} = H, \quad\Gamma_{rr}^{t} = \frac{H}{c^2}e^{2Ht}
\label{eqSitterChris}
\end{equation}
of the de Sitter space, Tina's four-acceleration $a^{\mu}$ for radial motion $(\vartheta, \varphi = \text{const})$ is given by
\begin{equation} 
  a^{t} = t'' + \frac{He^{2Ht}}{c^2}r'^2, \quad a^{r} = r'' + 2Hr't',
\label{eq:fouracc}
\end{equation}
and $a^{\vartheta} = a^{\varphi} = 0.$ Inserting Eq.~(\ref{eq:fouracc}) into Eq.~(\ref{eq:hypmotion}) and using the relations 
\begin{subequations}
 \begin{align}
  r'^2 &= c^2\left(t'^2-1\right)e^{-2Ht}, \label{eq:xssquared} \\
  \frac{r''}{r'} &= \frac{t't''}{t'^2-1} - Ht'. \label{eq:xslog} 
  \end{align}
  \label{eq:xs} 
\end{subequations}
which follow from a similar calculation as Eqs.~(\ref{eq:constraintflat1}) and (\ref{eq:constraintflat2}), Rindler derives the differential equation 
\begin{equation}
\begin{aligned}
 \left(\frac{\alpha}{c}\right)^2 = &\Bigg\{t'^2\left(t'^2-1\right)\left(\frac{t''}{t'^2-1} + H\right)^2 
\\&- \left[t''+H\left(t'^2-1\right)\right]^2\Bigg\}
\end{aligned}
\label{eq:DGLt}
\end{equation}
for $t'$, cf. Eq.~(\ref{eq:DiffEqFlat}). The substitution
\begin{equation}
 t' = \cosh(z)
\label{eq:substitution}
\end{equation}
simplifies Eq.~(\ref{eq:DGLt}) to 
\begin{equation}
 \left(\frac{\alpha}{c}\right)^2 = \left[z' + H\sinh(z)\right]^2.
\label{eq:DGLRindler}
\end{equation}
Integrating Eq.~(\ref{eq:DGLRindler}) and rearranging the terms yields
\begin{equation}
 \dfrac{S + \frac{\alpha}{c}\tanh\left(\dfrac{z}{2}\right)}{D - \frac{\alpha}{c}\tanh\left(\dfrac{z}{2}\right)} = Ae^{q\tau}.
 \label{eq:Int}
\end{equation}
When Tina leaves the origin from rest at time $\tau_{0}=0$, we have $\beta(0) = 0$ and, therefore, $t'(0) = \gamma(0)=1$, thus $z_{0}=0$. 
Hence, $A_{\tau_{0}}=S/D$. Now $t'$ can be calculated using the relation 
\begin{equation}
 \cosh(z) = \frac{1+\tanh^{2}(z/2)}{1-\tanh^2(z/2)}
\end{equation}
and Eq.~(\ref{eq:Int}) as
\begin{equation}
 t' = \dfrac{q(S^{2q\tau}+D)}{\Psi(\tau)}.
\label{eq:Rindlert}
\end{equation}
Integrating Eq.~(\ref{eq:Rindlert}) and adjusting the constant of integration such that $t(0)=0$ yields Eq.~(\ref{eq:tcomplete1}).
Using the positive root of Eq.~(\ref{eq:xssquared}) and the relations
\begin{subequations}
\begin{align}
 e^{-Ht(\tau)} &= \dfrac{2q^2e^{q\tau}}{\Psi(\tau)}, \label{eq:Expressions1a} \\
 \sqrt{t'^{2}-1} &= \frac{\alpha}{c}\frac{(e^{q\tau}-1)(Se^{q\tau} + D)}{\Psi(\tau)} \label{eq:Expressions1b}
\end{align}
\label{eq:Expressions1}
\end{subequations}
together with the condition $r(\tau=0)=0$, Rindler arrives at Eq.~(\ref{eq:xcomplete1}).

\section{Extension for a decelerating observer}\label{appsec:DecObs}
If Tina starts decelerating at $\tau=\tau_{1}$ by changing the proper acceleration via $\alpha\rightarrow -\alpha$, 
we must ensure that $t'$ is continuous at $\tau=\tau_{1}$ and that $r$ and $t$ are differentiable.
The condition of continuity for $t'$ is fulfilled if
\begin{equation}
 z_{1} = \text{acosh}\left[t'(\tau_{1})\right],
\end{equation}
cf. Eq.~(\ref{eq:substitution}).
Using the trigonometric relations
\begin{equation}
 \tanh(z/2) = \frac{e^{z}-1}{e^{z}+1}, \quad\text{acosh}(x) = \ln\left(x+\sqrt{x^{2}-1}\right),
\end{equation}
we obtain
\begin{equation}
\begin{aligned}
\tanh(z_{1}/2) &= \frac{t'(\tau_{1}) + \sqrt{t'(\tau_{1})^2-1}-1}{t'(\tau_{1}) + \sqrt{t'(\tau_{1})^2-1}+1} \\
		&= 1 - 2\frac{\Psi(\tau_{1})}{\Omega(\tau_{1})},
\end{aligned}
\label{eq:tanhnew}
\end{equation}
with $\Psi(\tau_1)$ and $\Omega(\tau_1)$ given in Eqs.~(\ref{eq:RindlerPsi}), (\ref{eq:omega1}).
Inserting Eq.~(\ref{eq:tanhnew}) into Eq.~(\ref{eq:Int}) yields the constant of integration $A_{\tau_1}$, cf. Eq.~(\ref{eq:Atau1}) and furthermore $t'$ for $\tau>\tau_{1}$.
The calculation for $\tau=\tau_3$ is performed in the same way and yields $B_{\tau_3}$, cf. Eq.~(\ref{eq:Btau2}).

Combining our result with Rindler's expressions for $\tau\leq\tau_{1}$ we obtain the piecewise definition of $t'$ in Eq.~(\ref{eq:tscomplete}).
Integrating Eqs.~(\ref{eq:tscomplete1}) and (\ref{eq:tscomplete2}) and choosing the additional constants of integration to fulfill continuity of $t(\tau)$
we obtain Eq.~(\ref{eq:tcomplete})
To calculate $r(\tau)$ we also use the positive root of Eq.~(\ref{eq:xssquared}) and the expressions analogue to Eq.~(\ref{eq:Expressions1}), choose the emerging constant of integration
properly and receive Eq.~(\ref{eq:xcomplete}).

\section{Maximum acceleration time\label{appsec:AccTime}}
In order to determine the maximum acceleration time $\tau_{1\text{max}}$ via
\begin{equation}
l(\tau_{\text{rest}}) = r_\infty,
\label{eq:taumaxdef}
\end{equation}
 we evaluate $r(\tau_{\text{rest}})$ and $e^{Ht(\tau_{\text{rest}})}$ with $\tau_{\text{rest}}$ given in Eq.~(\ref{eq:taurest}), and using Eqs.~(\ref{eq:tcomplete2}) and (\ref{eq:xcomplete2}), where
\begin{equation}
\tau_{\text{rest}} = \tau_{\text{rest}}(\tau_{1\text{max}}),
\end{equation}
 and obtain
\begin{equation}
\begin{aligned}
 e^{Ht(\tau_{\text{rest}})} &= A_{\tau_{1\text{max}}}\frac{D}{S}\frac{\Psi(\tau_{1\text{max}})}{\Psi_A(\tau_{1\text{max}})}, \\
 r(\tau_{\text{rest}}) &= \alpha\frac{\Psi_A(\tau_{1\text{max}})}{\Psi(\tau_{1\text{max}})}\frac{1}{HDA_{\tau_{1\text{max}}}} + \mathcal{K}(\tau_{1\text{max}}) + r_\infty.
\end{aligned}
\end{equation}
Multiplying these expressions, subtracting $r_\infty$ on both sides of Eq.~(\ref{eq:taumaxdef}), and taking out a factor $\Psi_A(\tau_{1\text{max}})^{-1}$
yields
\begin{equation}
 A_{\tau_{1\text{max}}}D\Psi(\tau_{1\text{max}})\left[\mathcal{K}(\tau_{1\text{max}}) + r_\infty\right] = 0.
\end{equation}
With $A_{\tau_1}, \Psi(\tau_1)\neq 0 $ we have
\begin{equation}
\mathcal{K}(\tau_{1\text{max}}) + r_\infty = 0
\label{eq:taumaxpoly}
\end{equation}
as defining equation for $\tau_{1\text{max}}$. 
To prove that $\tau_{1\text{max}}$ given in Eq.~(\ref{eq:taurest}) is a solution of this equation, we insert Eq.~(\ref{eq:taumax}) into Eq.~(\ref{eq:taumaxpoly}). With Eq.~(\ref{eq:KK})
 and the intermediate results
\begin{equation}
\begin{aligned}
\Psi(\tau_{1\text{max}}) &= \frac{q^2}{H}(3q+5H), \\
 A_{\tau_{1\text{max}}} &= 2\frac{H^2}{Hq+\frac{\alpha^2}{c^2}-H^2},
\end{aligned}
\label{eq:tau1maxintermediate}
\end{equation}
this can easily be shown.

\section{Maximum coordinate distance during a round trip\label{appsec:Rmax}}
As discussed in Sec.~\ref{sec:MaxAccTime}, a round trip is only possible for destinations with 
\begin{equation}
 r<r_{\text{max}} = r_\infty e^{-Ht(\tau_{\text{restmax}})}.
\end{equation}
To calculate $e^{-Ht(\tau_{\text{restmax}})}$, we use the results in Eq.~(\ref{eq:tau1maxintermediate}) and additionally
\begin{equation}
 \Psi_A(\tau_{\text{rest}}) = 2q^2,
\end{equation}
which is true for any $\tau_{\text{rest}}$ from Eq.~(\ref{eq:tau1max}) and
\begin{equation}
 \Psi(\tau_{\text{restmax}}) = \frac{A_{\tau_{1\text{max}}}^2D^2S}{H} + \frac{2}{H}D^2SA_{\tau_{1\text{max}}}-HD.
 \label{eq:rmaxintermediate}
\end{equation}
With Eqs.~(\ref{eq:tau1maxintermediate}), (\ref{eq:rmaxintermediate}) and (\ref{eq:tcomplete2}) we obtain
\begin{equation}
 e^{-Ht(\tau_{\text{restmax}})} = \frac{S}{2}\frac{2H^4-H^2\frac{\alpha^2}{c^2}-2Hq^3-3\frac{\alpha^4}{c^4}}{\left(3q+5H\right)\left(H^2-\frac{\alpha^2}{c^2}-Hq\right)q^2}.
\end{equation}


\end{document}